\documentclass[%
reprint,
 amsmath,amssymb,
 aps,
]{revtex4-1}

\usepackage{graphicx}
\usepackage{dcolumn}
\usepackage{bm}
\usepackage{siunitx}
\usepackage{subcaption}

\begin{document}

\preprint{APS/123-QED}

\title{Proof-of-principle direct measurement of Landau damping strength at the Large Hadron Collider with an anti-damper}

\author{Sergey~A.~Antipov}
    \email{Sergey.Antipov@cern.ch}

\author{N.~Biancacci}
\author{X.~Buffat}
\author{E.~M\'etral}
\author{N.~Mounet}
\author{D.~Valuch}
\affiliation{%
 CERN, Geneva, Switzerland
}%

\author{D. Amorim}
    \affiliation{CERN, Geneva, Switzerland}
    \affiliation{Synchrotron SOLEIL, Gif-sur-Yvette, France}

\author{A. Oeftiger}
    \affiliation{GSI, Darmstadt, Germany}
    \affiliation{CERN, Geneva, Switzerland}%

\date{\today}

\begin{abstract}

Landau damping is an essential mechanism for ensuring collective beam stability in particle accelerators. Precise knowledge of how strong Landau damping is, is key to making accurate predictions on beam stability for state-of-the-art high energy colliders. In this paper we demonstrate an experimental procedure that would allow quantifying the strength of Landau damping and the limits of beam stability using an active transverse feedback as a controllable source of beam coupling impedance. In a proof-of-principle test performed at the Large Hadron Collider stability diagrams for a range of Landau Octupole strengths have been measured. In the future, the procedure could become an accurate way of measuring stability diagrams throughout the machine cycle.

\begin{description}
\item[PACS numbers]
29.27.--a, 29.27.Bd

\end{description}
\end{abstract}

\maketitle


\section{\label{sec:LD}Landau Damping}

Landau damping is the damping of collective oscillation modes in collisionless plasmas, where particles interact via long-range forces. The damping arises through energy transfer from collective modes to the incoherent motion of resonant particles. It was predicted by Landau in 1946 \citep{Landau:1946jc} and observed by Malmberg and Wharton in 1964 \citep{Malmberg:1964}. A rigorous mathematical proof of the general nonlinear problem has been established relatively recently by Villani and Mouhot \citep{Villani:2011}.

In particle accelerators Landau damping is not a dissipative mechanism but a mechanism stabilizing the beam against an external excitation or beam induced wake fields. The damping arises from a natural spread of synchrotron or betatron frequencies of particles in the beam. The effect is important for stabilization of intense beams both in the longitudinal, as first suggested by Nielsen et al. \citep{Nielsen:1959}, and in the transverse degrees of freedom \citep{Lasslet:1965}. In synchrotrons Landau damping is an essential mechanism responsible for suppression of transverse head-tail intra-bunch modes that cannot be otherwise damped by a narrow-band feedback or mitigated by the choice of chromaticity. Nowadays precise knowledge of the strength of Landau damping is key to making accurate predictions on beam stability for high-energy colliders such as Large Hadron Collider (LHC) \cite{Bruning:2004ej,Gareyte:1997ds} or its High Luminosity upgrade \cite{Apollinari:2015bam} as well as future machines such as the Future Circular Collider with its hadron version (FCC-hh)~\cite{Benedikt:2018csr}. Also, intermediate-energy hadron machines running at high intensities such as the FAIR synchrotron SIS100 rely on Landau damping to suppress beam instabilities~\citep{bib:Kornilov_oct, bib:FAIR_TDR}.

A common technique of measuring Landau Damping is by means of Beam Transfer Function (BTF) \citep{bib:Handbook}, where the frequency dependence of the response to forced beam oscillations is used to quantify the Stability Diagram (SD) \citep{bib:Vaccaro_SD}. BTF has been successfully used to measure stability diagrams at GSI \citep{bib:Kornilov}, RHIC \citep{Luo:2018plt} and at injection energy in LHC \citep{bib:Tatiana}. The method has some limitations though: first, it might be challenging to maintain both good beam stability and high signal to noise ratio when driving the oscillation as seen at top energy in LHC \citep{bib:Claudia}. Second and most importantly, the measurement does not test the strength of the Landau damping itself, but the transfer function. Numerous approximations are usually made to obtain the SD from the BTF: the synchrotron frequency spread is neglected, the betatron frequency spread is assumed to be small, the beam response to an external excitation is assumed to be linear.

A new alternative approach for measuring the strength of Landau damping involves using the transverse feedback with a reverted polarity (anti-damper) to excite a collective mode in the beam. The anti-damper such acts as a controllable source of beam coupling impedance. By knowing the strength of the feedback excitation, and observing at which feedback gain the beam becomes unstable, one obtains a direct measurement of the strength of Landau damping in the synchrotron. Further, with an accurate control over the feedback phase one can explore the full complex plane of tune shift and growth rate. One can such derive the SD and compare with theoretical predictions. In this paper we describe a proof of principle test to measure the strength of Landau damping created by the LHC octupole system at 450 GeV injection energy and present its results.

\section{Measurement of Landau damping at LHC}

\subsection{\label{sec:Antidamper}Feedback as Controlled Impedance}

\begin{figure}[b]
  \centering
  \includegraphics[width=.75\linewidth]{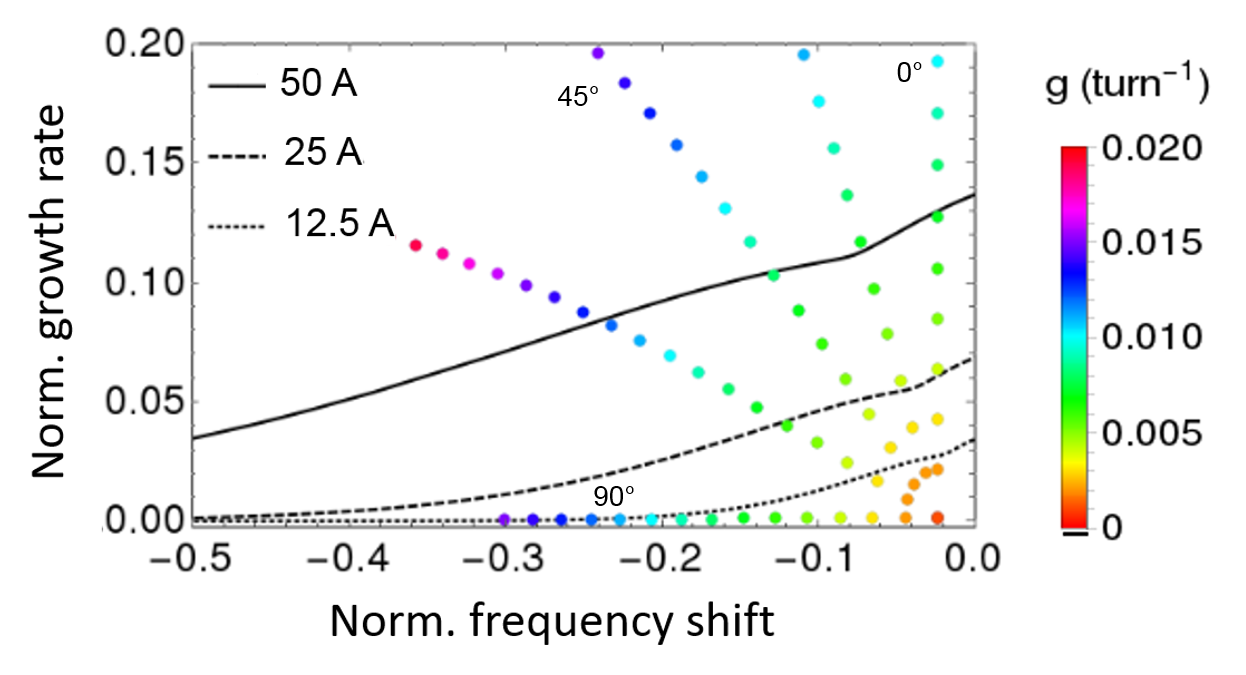}
  \caption{
    Controlling the gain and the phase of the feedback one explores the full relevant area of Stability Diagram in LHC. Real and imaginary mode frequency shifts are normalized by the synchrotron frequency $\omega_s$. SDs for a nominal $1.0~\mu$m emittance and quasi-parabolic beam distribution are shown in black.
    }
  \label{fig:modes}
\end{figure}

If the variation of the feedback's dynamic response over the bunch length can be neglected, i.e. it is `flat', it can be described as a constant wake force acting on the beam $W(z) = W_0 = const$. This is true e.g.\ for the LHC transverse feedback whose bandwidth goes up to 40 MHz or 1/10 of the Radio Frequency bucket width. This wake function corresponds to a $\delta$-function-like coupling impedance (for reference see, for example, \citep{Chao} or \citep{Burov:2013vya}):
\begin{equation}
    \centering
    Z_{d}(\omega) \sim i g \times e^{i\phi} \times \delta(\omega),
\end{equation}
where $g$ stands for feedback gain in inverse turns and $\phi$ for its phase: 0 indicates a resistive feedback (picking up on beam position) and 90 deg a reactive one (picking up on transverse beam momentum). A resistive feedback thus drives a coherent beam mode upwards in the diagram, driving it unstable, with an instability growth rate of $-g$ (Fig.~\ref{fig:modes}). Such a system has been proposed for the IOTA ring \citep{Antipov:2016ixg,bib:Antidamper}, where the researchers considered an anti-damper with $\phi = 0$.

\begin{figure}
  \centering
  \includegraphics[width= 2 in]{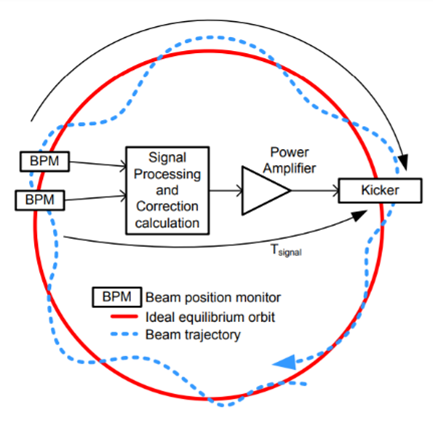}
  \caption{
    LHC feedback system uses two pickups per plane allowing acting independently on both beam transverse position and its transverse momentum.
    }
  \label{fig:LHC_feedback}
\end{figure}

A realistic impedance of various accelerator components ranges from inductive impedance of high-Q RF modes, to broadband imaginary impedance of bellows and tapers. These impedances can be modelled by different phases of the feedback: from 0 for a purely imaginary tune shift to 90 deg for a purely real one. In practice, a variation of the phase is convenient to achieve with two feedback pick-ups: one picking up on the beam momentum and the other on its position. The LHC transverse feedback system features two pick-ups  (Fig.~\ref{fig:LHC_feedback}) that allows producing an arbitrary complex gain $g \times e^{i\phi}$ by adjusting phase delay between them and their gain~\citep{Butterworth:2016fum}.

\subsection{Measurement setup}

A proof-of-principle test has been performed at nominal LHC injection settings with an injection probe beam, i.e.\ a single bunch of $0.5\times 10^{10}$ p in 1 $\mu$m normalized rms emittance (Table \ref{table:1}). At these parameters the azimuthally dipolar head-tail mode typically dominates the landscape with higher order modes featuring much weaker growth rates. Working with probe beams allows minimizing mode shift from beam coupling impedance as well as reducing space charge effects (discussed below in more detail). At the same time, working at injection energy allows to imply large octupolar detuning in order to stabilize the beam without the feedback, which is very useful when setting up the measurement. Usage of the nominal settings has been imperative to ensure rapid machine set-up and re-injection, which is needed for repetitive measurements as well as precise control of the optics.

\begin{table}[b]
\centering
\caption{Key parameters used for the study}
\begin{tabular}{l l} 
 \hline\hline
 Parameter & Value \\ [0.5ex] 
 \hline
 Beam energy & 450 GeV \\ 
 Beam intensity & $0.5\times 10^{10}$ ppb \\
 Number of bunches & 1 \\
 Norm.~tr.~emittance, rms & 1.0-1.1 $\mu$m \\
 Bunch length, $4\sigma_{rms}$ & 1 ns \\
 Coupling, $|C^-|$ & 0.001 \\
 RF voltage & 6 MV \\
 Tunes: x, y, z & 0.275, 0.295, 0.005 \\
 Chromaticity, $Q'$ & 14 \\
 Synch. freq., $\omega_s$ & 0.03 rad$^{-1}$\\ 
 SC tune shift & $\mathcal{O}\bigl(10^{-4}\bigr)$ \\
 [1ex] 
 \hline\hline
\end{tabular}
\label{table:1}
\end{table}

In LHC the betatron frequency spread required to produce the Landau damping is largely generated by a dedicated system of of 84 focusing and 84 defocusing 30 cm long superconducting octupoles~\citep{Bruning:2004ej}. First, the feedback was calibrated and qualified to act as an effective impedance, and then it was used to measure Landau damping produced by machine nonlinearities and by several configurations of octupole current: +11, +17, and -11~A and two values of chromaticity $Q' = +3, +14$.

\subsection{Measurement procedure}
\subsubsection{Feedback calibration}

\begin{figure}[h]
  \centering
  \includegraphics[width = 3.125in]{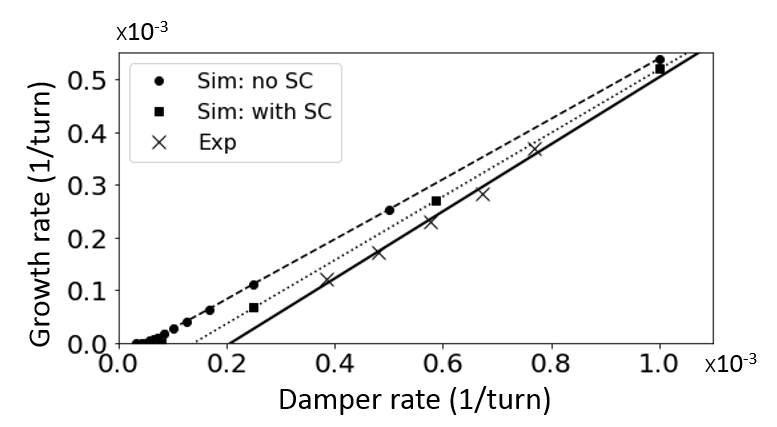}
  \caption{
    Instability growth rate scales linearly with the damper gain, allowing to calibrate the feedback strength. The non-zero gain required to start an instability is caused by natural nonlinearities of the machine. Overall, experimental data (crosses and the solid line) is in good agreement with numerical simulations (squares and the dotted line). Numerical simulations performed with space charge show a greater amount of feedback gain required to destabilize the beam than in the no-space-charge case (circles and the dashed line), emphasizing the importance of accounting for the space charge interaction at the LHC injection energy, $E = 450$~GeV.
	}
  \label{fig:damper_calibration}
\end{figure}

\begin{figure}[b]
  \centering
  \includegraphics[width = 3.5in]{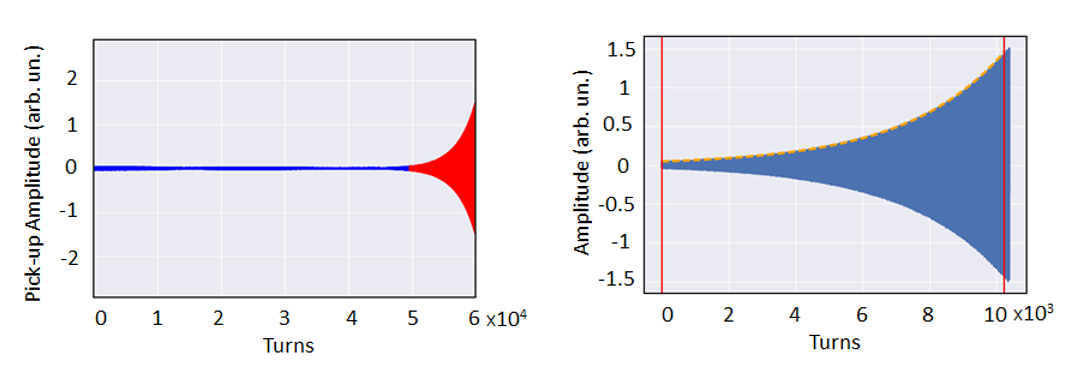}
  \caption{
    An example of an instability observed during the feedback calibration process. Left -- the full 64000 turn acquisition of the center of mass position, the unstable area is highlighted in red. Right -- zoom-in of the instability. Dashed yellow line represents an exponential fit of the data.
    }
  \label{fig:data_exapmle}
\end{figure}

In order to ensure an independent control over both the feedback gain and phase the system was calibrated with no octupole current at three anti-damper phases: 0, 45 and 65.7 deg. The resulting dependence of the instability growth rate on the feedback gain was found to be linear, as expected (Fig. \ref{fig:damper_calibration}). Also, the growth rate slope reduces gradually with the phase, as expected. The magnitude of the slope yields the calibration factor for the feedback gain (i.e. a setting of $x$ units drives an instability with an exponential rise time of $y$ turns) for the following measurements. An example of raw data is shown in Fig.~\ref{fig:data_exapmle}.

The calibration has also been verified via a tune shift measurement at a 45~deg feedback phase. If the feedback acts as an impedance whose strength increases linearly with the feedback gain $g$, then the (real) dipolar tune should shift linearly with $g$ -- which could be confirmed. The corresponding slope yields $13\times 10^{-3}$. From the above growth rate calibration, which corresponds to the imaginary tune shift, one would infer a slope of $\frac{1}{2 \pi} 63.9\times 10^{-3} = 11\times 10^{-3}$. A small discrepancy can be explained by the uncertainty of tune shift determination due to a large jitter present in the data when the beam became unstable. An uncertainty in feedback phase can also contribute to the difference but, based on a post-processing of the signals measured by the feedback, the error of phase setting should not have exceeded $\sim 1$~deg.

\subsubsection{Stability Diagram scans}

\begin{center}
    \begin{figure}
      \centering
      \includegraphics[width = 7in]{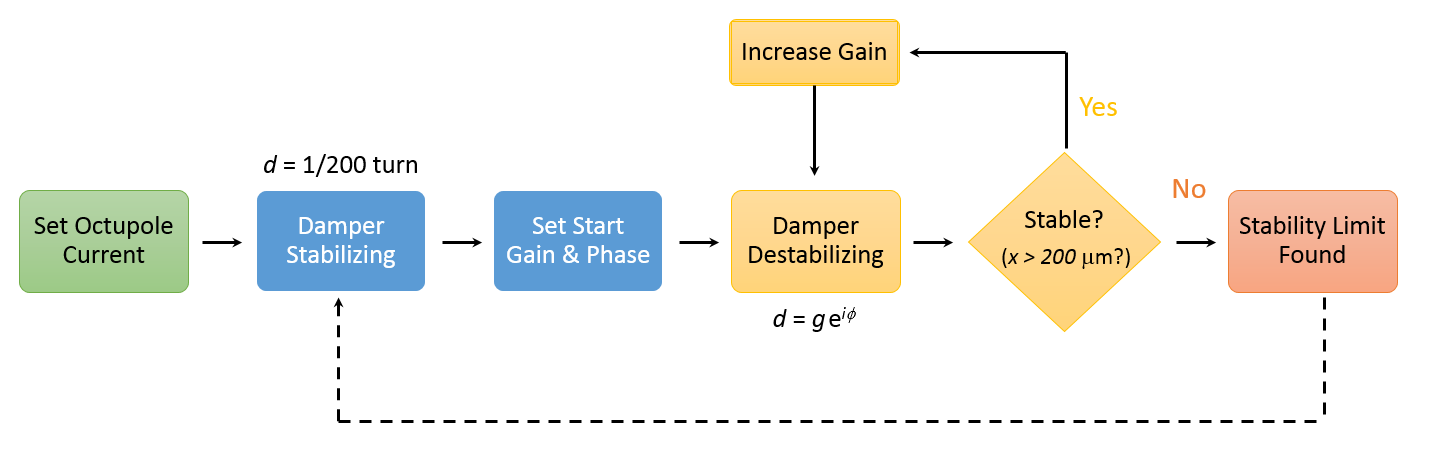}
      \caption{
        Procedure for measuring stability diagrams: feedback gain is increased at a fixed phase until a threshold amplitude is exceeded then the feedback is reverted back to stabilising.
        }
      \label{fig:Proc}
    \end{figure}
\end{center}

After calibrating the feedback we performed a series of measurements at different octupole settings. At each setting the feedback gain was gradually increased in small steps until reaching the limit of stability. At this point the feedback phase was increased -- as the stability diagram contour should have an increasing distance to the origin at increasing phase, the beam should return to stable conditions at the higher phase. This procedure has been repeated for 5 different phases between 0 and 90 deg. At each step the feedback gain was kept constant for about 30~sec, which should have excluded potential impact of latency effects. This time window has been chosen following a recent study, where latency effects could be excluded in single bunch octupole threshold measurements with sufficiently short, about 1~min steps \citep{Xavier:2019}. In our experiment, an instability was declared as soon as the beam centroid excursion from the reference orbit exceeded 200~$\mu$m -- a value comparable to the rms transverse size of the beam. In this case the feedback was automatically switched back to a resistive stabilizing mode with a damping time of 200 turns~(Fig.~\ref{fig:Proc}).

\onecolumngrid
\twocolumngrid

Normally beam emittance was not affected by the measurement, according to beam synchrotron radiation monitor. Therefore, in order to save time, the beam was re-injected only if a blow-up had been observed and the emittance increased by over 10\%. The downside of this approach is that the distribution tails, which have a large tune shift and thus play a large role in Landau damping, might have been affected by the previous measurements. While no systematic study of the effect was attempted during this proof-of-principle test, several data points were measured twice to check the reproducibility of the results. The results with an `old' and with a `fresh' turned out to be in good, 10\% or better agreement.

\section{Results and discussion}

In this section we discuss the results of measurements of stability diagrams at different octupole and chromaticity settings at injection energy, provide an estimate of its fill-to-fill variation arising from slight differences in beam parameters, and discuss the phenomena that might have affected the measurement.

\subsection{Landau damping by natural nonlinearities}

\begin{figure}[h!]
  \centering
  \includegraphics[width=\linewidth]{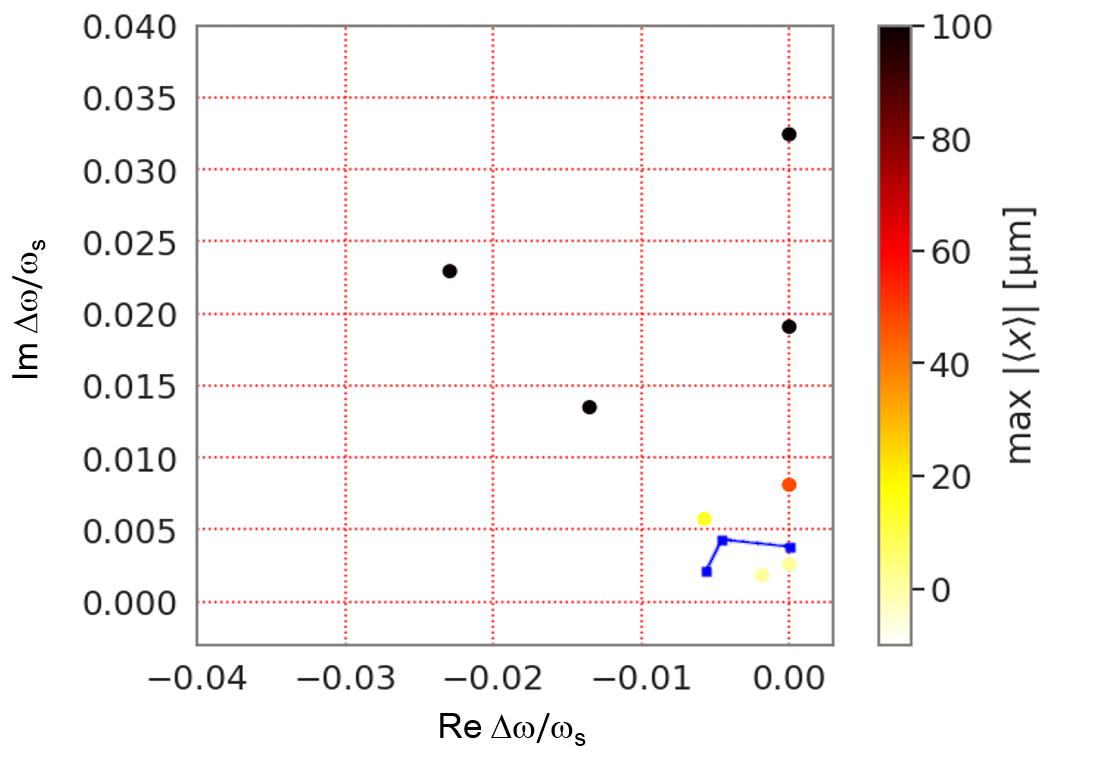}
  \caption{
    Stability region in complex damper configuration plane, simulated for the assumed natural nonlinearities in LHC ($E = 450$~GeV, effective octupole current $-2.5$~A). Blue line shows the limit of stability found in the measurement.
   	}
  \label{fig:PyHHT_nat_nonl}
\end{figure}

The calibration plot in Fig.~\ref{fig:damper_calibration} gives an insight into the strength of Landau damping by natural nonlinearities of the lattice, which can be seen as the intersection of the lines with the horizontal axis. These are the points where the beam as a dynamical system is exactly at the limit of stability. For a resistive feedback, corresponding to purely real impedance, an excitation rate of about $2.5\times 10^{-4} = 1/4000~{\text{turn}}^{-1}$ is needed to destabilize the beam. The blue line in Fig.~\ref{fig:PyHHT_nat_nonl} shows the corresponding stability diagram found by extrapolating the measurement data points at different phases to Im~$\Delta Q = 0$. It resembles the shape found in the numerical simulation, although suggesting a larger stable area. A part of the discrepancy might come from the inability to trigger on very small orbit excursions and thus observe instabilities with very small growth rates in the experiment.


Injection-to-injection spread of the strength of Landau damping, measured over 5 consecutive injections at 11 A and 0 deg phase (resistive anti-damper) turned out to be rather small - around 7\%, indicating a quite good reproducibility of beam distribution. The 7\% value gives lower limit of the systematic uncertainty in all the subsequent measurements. Depending on the direction of the shift the modes would probe different parts of the octupole SD: for the imaginary shift it would be the center that is nearly independent of the beam distribution or the octupole polarity, whereas for the real shift – it would be the tail of the diagram that drastically depends on the beam parameters (i.e. emittance, intensity, bunch profile, etc.), which all vary slightly fill-to-fill. The uncertainty of the SD tails can therefore be significantly larger than that of for the central peak.

\subsection{Landau damping by octupoles}

\begin{figure}
  \centering
  \includegraphics[width = 3in]{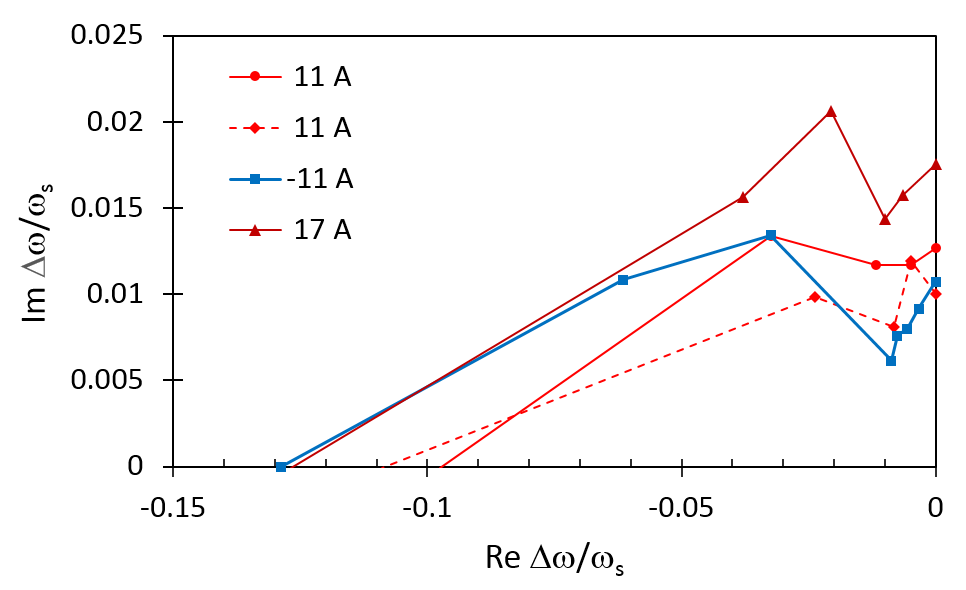}
  \caption{
    The height of the diagrams scales linearly with the octupole current with the negative octupole polarity providing around 30\% larger coverage of negative mode frequency shifts, which are relevant for coherent beam stability. LHC stability diagrams were measured at 450~GeV in the horizontal plane, solid lines -- $Q' = 14$, dashed line -- $Q' = 3$. Real and imaginary tune shifts are normalized by the synchrotron frequency.
	}
  \label{fig:SD_current_comp}
\end{figure}

The shape of the measured SDs at 11, 17, and -11~A qualitatively matches the expectations from a simple linear SD theory. Both the height and the width scale with the octupole current, with the SD for 17~A being around 50\% higher than that for 11~A, as expected (Fig.~\ref{fig:SD_current_comp}). The second measurement for 11~A current made at a lower chromaticity of 3 units matches within $10-20$ percent the first one performed at 14 units. The negative octupole polarity offers around 30\% greater coverage of the negative tune shifts, also in good qualitative agreement with what one would expect from a diagram of the octupole tune spread (Fig.~\ref{fig:coupling_tune}). This illustrates why the negative polarity is preferred to suppress impedance-driven instabilities in LHC that feature negative mode frequency shifts. The exact magnitude of the gain should depend on the details of the beam distribution.  

\subsection{Factors affecting the measurement}
\subsubsection{Impact of lattice nonlinearities}

The major part of lattice nonlinearities in LHC is created on purpose by its octupole system, generating an rms detuning of the order of $10^{-4}$ at injection in normal operation. Yet there are several other sources of nonlinearities affecting Landau damping at injection energy. The first one is the hysteresis effect of the octupole correctors affecting their magnetic field. It is responsible for a systematically larger measured $Q''$ in the machine in comparison to prediction \citep{Ewen:Hyst}. Another issue identified in \citep{Maclean:2014qna} is the large working point-dependent nonlinear amplitude detuning in the plane where one approaches the coupling resonance. This does not affect the present measurement as the real tune shifts during SD scans did not exceed 0.001. Finally, recent findings indicate a potential feed-down from decapole correctors through a systematic misalignment of about 0.25~mm \citep{Ewen:FiDel_2019}. Overall, these nonlinear effects generate a linear detuning with amplitude that is roughly equivalent to about $-2.5$~A of octupole detuning, thus providing $\sim 10^{-5}$ rms tune spread. The natural nonlinearities can therefore be largely omitted from the analysis, except for the case of vanishing Landau Octupole current.

\subsubsection{Impact of linear coupling}

\begin{figure}
  \centering
  \includegraphics[width=.75\linewidth]{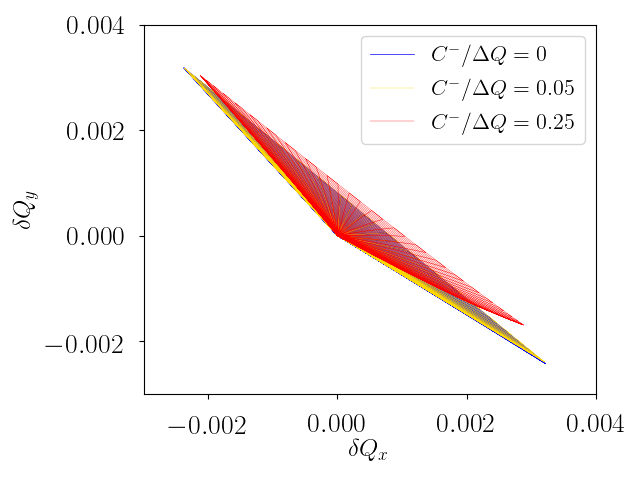}
  \caption{
    Tune footprint expected to be created by an $I_{oct} = 11$~A octupole current at the injection energy in LHC. The footprint can be modified slightly by Linear coupling. Tune footprints computed using numerical tracking in the MAD-X code~\citep{bib:MAD-X}.
    }
  \label{fig:coupling_tune}
\end{figure}

\begin{figure}
  \centering
  \includegraphics[width=.75\linewidth]{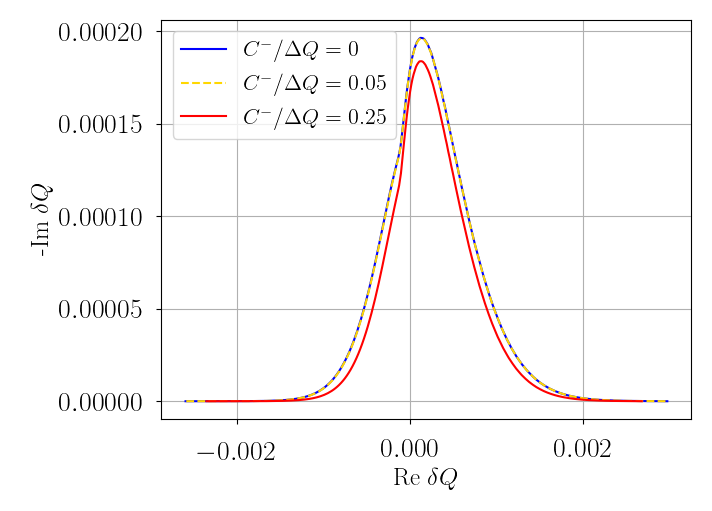}
  \caption{
    Linear coupling reduces the height of the stability diagram.  $I_{oct} = 11$~A, $E = 450$~GeV.
	}
  \label{fig:coupling_SD}
\end{figure}

Another significant effect is the linear coupling that can change amplitude detuning from Landau octupoles, reducing the footprint locally, but leading as well to a large second order amplitude detuning \citep{Maclean:2014qna, Maclean:2017axj}. It is important to note that the value of $|C^-|$, i.e. the global coupling defined as the closest tune approach, may change if uncorrected by as much as $5\times 10^{-3}$ over several hours at injection due to alignment errors of sextupole spool pieces \citep{Tobias:Priv_Comm}. While the coupling had been corrected down to a sufficiently low value in the beginning of the test $|C^-| = 0.001$, it may have drifted away from the initial value over time, which would shrink the octupole tune footprint (Fig.~\ref{fig:coupling_tune}) and result in a slightly smaller the SD for larger $|C^-|$ values~\citep{bib:Lee_C}. Figure \ref{fig:coupling_SD} depicts the effect of linear coupling at the octupole strength. At 11~A the reduction of the stability diagram does not exceed $10~\%$.

\subsubsection{Impact of impedance}
\begin{figure}[h!]
  \centering
  \includegraphics[width=.95\linewidth]{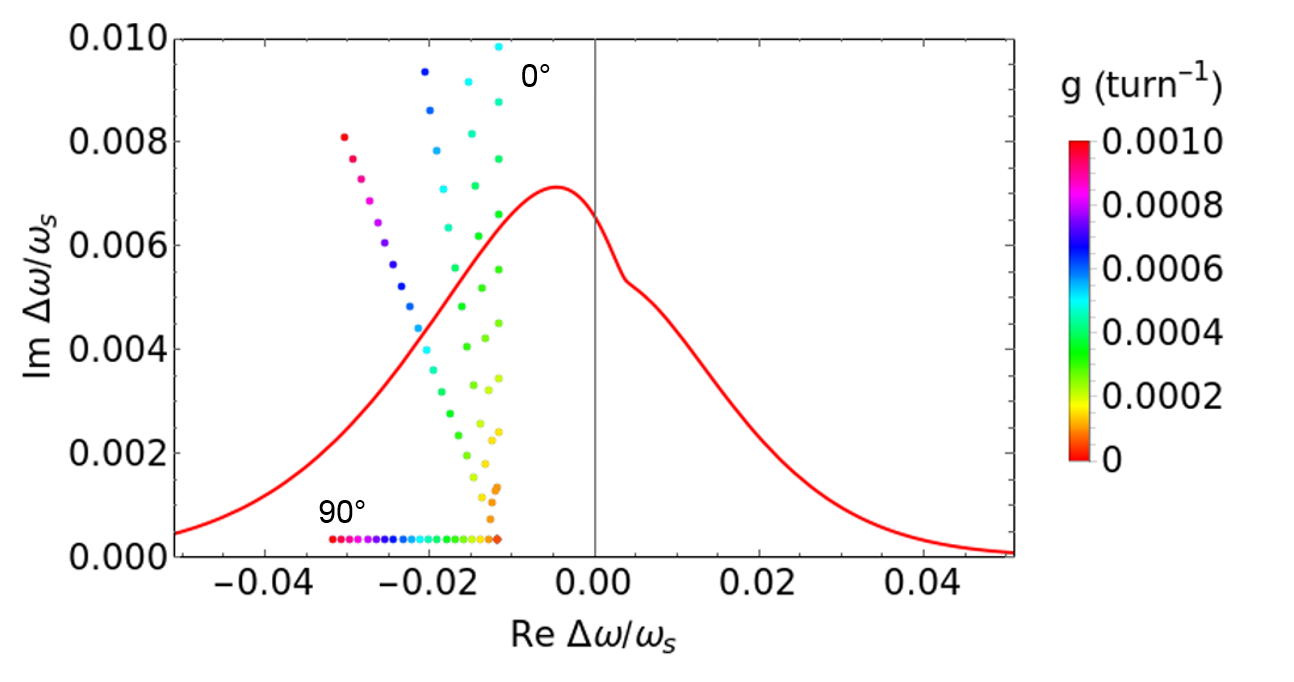}
  \includegraphics[width=.95\linewidth]{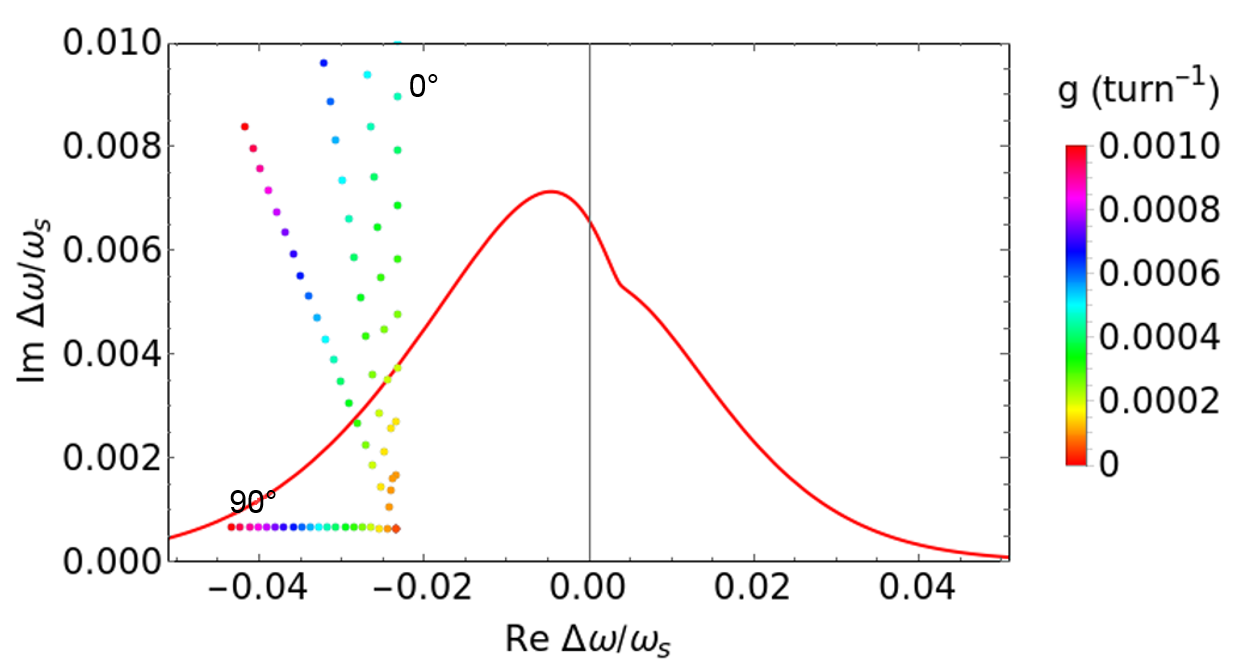}
  \caption{
    Machine impedance creates a negative tune shift, affecting the position of stability diagram crossing when a destabilizing feedback is applied. Top -- nominal machine impedance; bottom -- double machine impedance. The stability diagram, depicted by a red line corresponds to a Gaussian beam with 1~$\mu$m rms normalized emittance and 2.5~A positive octupole current. Various feedback gains for 5 equidistant phases between 0 and 90 deg are shown as coloured dots.
	}
  \label{fig:impedance}
\end{figure}

The complex frequency shift of a mode is also affected by the machine's impedance, and therefore the latter has to be taken into account. Underestimating the impedance might lead to a dramatic miscalculation of the octupole threshold as shown in Fig.~\ref{fig:impedance}. With two times the impedance the dipolar mode obtains a significant negative tune shift. If one excites it with a resistive feedback it crosses the stability diagram at a different location, closer to the tail of the diagram, at a factor two lower feedback gain. If one then uses this lower gain to deduce the octupole threshold without considering the mode shift produced by the impedance, one might underestimate the threshold by about a factor two.

\subsubsection{Impact of space charge}

Although space charge on its own does not provide Landau damping for the rigid dipole mode, as pointed out by M\"ohl \citep{Mohl:1995gb}, it does modify the SD produced by lattice nonlinearities. In general,  an interplay of octupole detuning and nonlinear space charge may be important as observed in particle tracking simulations \citep{Kornilov:2008zz}. When the strength of space charge detuning is small relatively to other sources its impact can be estimated analytically in a simple model \citep{Metral:2004mc}, assuming a quasi-parabolic transverse distribution, coasting beam, and a linear space charge detuning (the model can be extended to bunched beams \citep{Ng:2008zzb}, although the impact of the bunching is minor). Depending on the strength of space charge, it leads to a negative real tune shift of the SD maximum, a widening of the diagram, and a slight reduction of its height. For the studied parameters, the impact of space charge should be relatively weak providing a shift of the SD of around $\Delta_0 = 10^{-4}$.

Since the space charge, even as weak as $\sim 0.1 Q_s$, affects Landau damping and, most importantly the feedback calibration (Fig.~\ref{fig:damper_calibration}), a quantitative analysis of the observed stability diagrams has to be done including it in the picture. Currently, to the authors' best knowledge, there are no satisfactory analytical models of Landau damping of bunched beams with space charge. One has to, at this stage, rely on computationally demanding macroparticle simulations. Direct measurements using an anti-damper can therefore be of great value for benchmarking such simulations. 

Several logical steps can be undertaken at LHC to investigate Landau damping with space charge further. First, after demonstrating sufficient safety for the machine, bunch intensity can be increased up to $\sim 10^{11}$~p or $\Delta Q_{SC} \sim Q_s$. This allows investigating how an increasing space charge force affects Landau damping by octupoles. Second, a test at the top energy of 7~TeV flat-top where space charge interaction is negligible and lattice nonlinearities are normally well controlled would allow benchmarking existing models used for stability prediction for LHC and its upgrades.

\subsection{Numerical Tracking Simulation}

As was shown above, the space charge interaction, in a linear model, can significantly shift the SD towards negative frequencies destabilising the beam where it otherwise should have been stable. To investigate the issue further we performed macro-particle simulations in the PyHEADTAIL macroparticle tracking code, which performs symplectic tracking in 6D \citep{bib:Li, bib:2016grn}. The tracking utilizes a smooth optics approximation and a drift/kick model, treating the accelerator ring as a collection of interaction points connected by ring segments where the beam is transported via linear transfer maps. Collective effects, arising from impedance, space charge, or external feedback are applied at the interaction points where the beam is longitudinally divided into a set of slices via a 1D particle-in-cell (PIC) algorithm. The transport maps take into account local Twiss parameters and dispersion at each interaction point but non-linearities such as chromaticity are included indirectly as an additional detuning applied to each individual macroparticle.

The numerical model included a $-2.5$~A equivalent octupole linear amplitude detuning, and nonlinear longitudinal motion but not the linear coupling effects. Without space charge, \SI{1d6}{macro}-particles have been tracked for \SI{1d6}{turns}. Simulations including self-consistent space charge (via a 2.5D slice-by-slice PIC algorithm) are based on \SI{3d6}{macro}-particles being tracked during \SI{60d3}{turns}.

\begin{table}[h!]
\centering
\caption{Key simulation parameters in addition to Table~\ref{table:1}}
\begin{tabular}{l l} 
 \hline\hline
 Parameter & Value \\ [0.5ex] 
 \hline
 Chromaticity & $Q'_{x,y} = 15$ \\
 Transverse tunes & $(Q_x, Q_y) = (64.28, 59.31)$ \\
 Synchrotron tune & $Q_s = \SI{4.9d-3}{}$ \\
 [1ex] 
 \hline\hline
\end{tabular}
\label{table:2}
\end{table}



\section{Conclusion and Outlook}

In this proof-of-principle test we have demonstrated that the active feedback system can be used as a source of controlled impedance to probe the strength of Landau damping. The experiment has been carried out in LHC at injection energy of 450~GeV with single low intensity bunches. First, the active feedback system has been calibrated to create an arbitrary complex tune shift. Both tune shift and instability growth rate have been demonstrated to increase linearly with the feedback gain, as expected. Then, the feedback has been utilized to directly measure the strength of Landau damping by gradually increasing its gain until a transverse activity is observed. The possibility of exploring the Stability Diagram by changing the damper phase has also been demonstrated. The results are in good qualitative agreement with the theoretical SD predictions. A detailed quantitative analysis (in particular, tracking simulations with space charge) is required to include effects of lattice nonlinearities and coherent effects in the picture. 

A good control and an accurate knowledge of natural machine nonlinearities seems important for being able to benchmark the results against Stability Diagram predictions. With a beam-based nonlinear correction one should be able to keep the lattice nonlinearities at injection at a sub 1~A level, as seen in 2016-17 \citep{Ewen:Review}. 

The technique has a potential to become a fast non-destructive tool for measuring the strength of Landau damping throughout the accelerator cycle. In LHC it would be well suited for studies at the top energy, where the constraints arising from Landau damping are the tightest and the effect of space charge is negligible. In order to explore this potential, further studies including the top energy of 7~TeV are required after the Long Shutdown.

\section*{Acknowledgments}
The authors would like to thank Alexey Burov (FNAL) for sharing his ideas on the use of a transverse feedback as a source of controlled impedance which triggered this research. The results presented in this paper have been obtained within tight time constraints during a study done in parallel with several others. We would like to express our gratitude to the Operations Team and especially Francesco Velotti for a fruitful and productive collaboration that helped sketching a realistic plan and ultimately successfully carry out tests. We shall also acknowledge Lukas Malina for his insightful tips on turn-by-turn data processing and sharing his Python scripts that were used in the present analysis, Gianluigi Arduini for his comments, Ewen Maclean and Tobias Persson for discussion on the sources of lattice nonlinearities in LHC.

\end{document}